\documentclass[10pt]{article}
\usepackage{fancyhdr}
\usepackage{amsmath}
\usepackage{amsthm}
\usepackage{amsfonts}
\usepackage{siunitx}
\usepackage{tikz}
\usepackage[plain]{algorithm}
\usepackage{algpseudocode}
\usepackage{multirow}
\usepackage{booktabs}
\usepackage{graphicx}
\usepackage{subfigure}
\usepackage{xcolor}

\usepackage{longtable}
\usepackage{bm}
\usepackage{mathtools}
\usepackage{amssymb}
\usepackage[labelfont=bf]{caption}
\usepackage{capt-of}
\usepackage{mciteplus}
\usepackage{cite}

\usepackage{mathrsfs}
\usepackage[title,titletoc,toc]{appendix}
\usepackage{xr}
\usepackage{parskip}
\usepackage{soul}
\usepackage{textcomp}
\usepackage[colaction]{multicol}
\usepackage[switch]{lineno}
\usepackage{lipsum}
\usepackage{etoolbox}
\usepackage{array}
\usepackage{tablefootnote}
\usepackage{ragged2e}
\newcolumntype{C}[1]{>{\centering\arraybackslash}p{#1}}
\usetikzlibrary{automata,positioning}
\usepackage{adjustbox}
\usepackage{makecell}
\usepackage{pdflscape}
\usepackage{tabularx}

\usepackage{setspace}
\usepackage{url}
\usepackage[utf8]{inputenc}
\usepackage[T1]{fontenc}
\usepackage{textgreek}

\usepackage[colorlinks=true,linkcolor=black,anchorcolor=black,citecolor=black,urlcolor=blue]{hyperref}

\newcommand{\DynFPs}{\textnormal{Dyn\_FPs}}
\newcommand{\TopFPs}{\textnormal{Top\_FPs}}
\newcommand{\GeoFPs}{\textnormal{Geo\_FPs}}

\theoremstyle{plain}

\theoremstyle{definition}

\newcommand{\keywords}[1]{\noindent\textbf{Keywords:} #1\par\vspace{0.5em}}



\begin{document}

\title{Chaotic Dynamics-Regulated Topological Learning for Patient-Specific Preictal State Identification} 

\author{Zihan Wang$^{1}$, Daixin Li$^{1}$, Guilin Wang$^{1}$, 
Mushal Zia$^2$, Xiaoqi Wei$^2$, \\ Xiang Xiang Wang$^2$, and Jian Jiang$^{1,2}$\footnote{Corresponding author. Email: jjiang@wtu.edu.cn} \\
$^{1}$Research Center of Nonlinear Science, School of Mathematics and Statistics, \\ Wuhan Textile University, Wuhan, 430200, P. R. China \\
$^{2}$Department of Mathematics, Michigan State University,\\ East Lansing, Michigan 48824, USA }
\date{}

\maketitle

\begin{abstract}
Epileptic seizures arise from complex, nonlinear interactions within brain networks, yet reliable electroencephalographic (EEG) prediction remains challenging due to the nonstationary and heterogeneous nature of neural dynamics. Existing methods typically analyze EEG data as static or weakly time-dependent snapshots. Consequently, they overlook the intrinsic dynamics and lack the geometric sensitivity required to capture the hierarchical, localized evolution of the epileptogenic zone. To address these limitations, we propose an offline, patient-specific evaluation of chaotic dynamics-regulated topological learning (CDRTL) for distinguishing preictal from interictal EEG states. This framework unifies chaotic dynamics, multiscale algebraic topology, and local network differentiation. Specifically, we first partition EEG signals into discrete functional subnets based on correlation strengths, effectively capturing the multi-scale connectivity of the brain. By modeling each node as a Lorenz oscillator, we embed the underlying chaotic signal dynamics into the network architecture. We then apply the persistent Laplacian to simultaneously extract topological invariants and geometric shape evolution through harmonic and non-harmonic spectral analysis. Additionally, we implement a node-removal topological differentiation strategy to isolate specific localized neural contributions. Our framework was evaluated on the CHB-MIT database using balanced preictal and interictal labels and stratified channel-level cross-validation performed independently within each patient. The results support offline discrimination of preictal and interictal channel-level nodes within fixed patient-specific networks. Because representations are constructed from the complete network, including held-out unlabeled nodes, before cross-validation, the reported performance is specific to this transductive setting and does not establish generalization to unseen EEG windows, seizures, or patients.\end{abstract}


\newpage

\section{Introduction} \label{sec1}

Epilepsy is a chronic neurological disorder characterized by recurrent, unprovoked seizures arising from abnormal, hypersynchronous neuronal activity, and it imposes substantial cognitive, psychological, and socioeconomic burdens on patients. Electroencephalography (EEG) remains the main noninvasive modality for monitoring brain activity and provides critical information on the temporal evolution of neural dynamics. In particular, distinguishing preictal states from interictal periods is of great clinical importance \cite{slama2025comprehensive}. However, EEG signals are inherently nonlinear, nonstationary, and often contaminated by noise, which makes reliable preictal state identification in seizure prediction a challenging task. These challenges have driven the development of computational approaches that have aimed at extracting informative features from complex EEG data.

Recent advances in machine learning (ML) have significantly reshaped the methodologies of preictal state identification. Early approaches relied on handcrafted features derived from time and frequency domains, such as statistical moments, wavelet coefficients, and decomposition-based features, which were subsequently processed using classical classifiers such as support vector machines (SVM) and random forests (RF) \cite{ALICKOVIC201894,usman2017epileptic,perez2022epileptic}. More recently, deep learning architectures, including convolutional and recurrent neural networks, have demonstrated strong performance by automatically learning hierarchical representations from raw EEG signals \cite{wu2025review, zhang2024review,saadoon2025machine}. In parallel, graph-based methods have been introduced to capture functional connectivity between EEG channels, modeling the brain as a networked system \cite{li2019transition}. Complementing these developments, topological data analysis (TDA) has emerged as a robust framework for extracting scale-invariant structural features from complex data. Persistent homology (PH), a central tool in TDA, characterizes the birth and death of topological features in multiscale filtrations and has been successfully applied to EEG analysis, revealing structural patterns not accessible by conventional feature engineering \cite{wee2025review,su2026topological,jiang2025machine}. Previous studies have shown that PH-derived descriptors, such as persistence landscapes and entropy, can improve classification performance and provide insight into brain network organization \cite{wang2018topological,prantzalos2023matilda}.

Despite these advances, several limitations remain. Many existing approaches treat EEG signals as static observations or as sequences of weakly dependent temporal segments, thereby neglecting the intrinsic dynamical structure governing neuronal activity. Seizure generation is inherently a dynamical process that involves nonlinear transitions toward synchronized states, often associated with chaotic behavior and critical phase transitions \cite{tan2025automatic}. However, current TDA-based methods typically operate on static point clouds or networks, failing to incorporate the temporal evolution and underlying physical mechanisms of brain dynamics. Additionally, PH captures topological invariants up to homotopy equivalence but lacks sensitivity to geometric properties and their evolution. As a result, it can identify the presence of features such as loops or connected components, but it cannot characterize their geometric configuration, density, or temporal deformation \cite{su2025topological,wee2025review,wei2025persistent2}. This limitation is particularly restrictive in EEG analysis, where subtle changes in synchronization patterns and spatial organization play a crucial role in distinguishing preictal from interictal states. Furthermore, PH primarily provides global summaries of the network, with limited capacity to resolve local contributions associated with epileptogenic regions \cite{wei2025persistent}. These observations highlight the need for a unified framework that integrates dynamical modeling, topological invariants, and geometric structure.

To address these challenges, we propose a chaotic dynamics-regulated topological learning (CDRTL) framework and evaluate it in an offline, patient-specific setting for identifying preictal EEG patterns associated with upcoming seizures. Prior to network construction, pairwise Pearson correlation coefficients are computed between EEG channels and used to partition the data into ten sub-networks according to interaction strength. This functional differentiation captures the multiscale organization of brain connectivity, reflecting strong, intermediate, and weak coupling regimes associated with epileptic dynamics. In particular, strongly connected regions correspond to hub-like structures near the epileptogenic zone \cite{hadjiabadi_maximally_2021}, intermediate connections form transition regions characterized by competing excitatory and inhibitory processes \cite{moosavi_criticality_2023}, and weak connections represent long-range communication that becomes attenuated during the preictal phase \cite{di_giacomo_static_2026}.

Building upon this multiscale representation, we employ the persistent Laplacian (PL) framework to extract both topological invariants and geometric evolution features through harmonic and non-harmonic spectral analysis \cite{wang2020persistent}. To further incorporate intrinsic dynamics, each node is modeled as a nonlinear oscillator governed by the Lorenz equations, resulting in a coupled dynamical system that captures both local signal behavior and global interactions. From this system, we derive complementary feature sets, including PL-based spectral features and dynamical features obtained from the temporal trajectories of the oscillators. These features are fused into a global representation of each EEG channel. To capture node-level contributions, a topological differentiation strategy is introduced \cite{zhang2025meta}, in which the global characteristic is recomputed after removing each node, and the resulting difference vector is used to characterize its influence \cite{chen2023path}. The multiscale features obtained from all sub-networks are concatenated and used for classification. The proposed CDRTL framework is evaluated on the CHB-MIT dataset, demonstrating that the integration of topology, geometry, and nonlinear dynamics provides an expressive, physics-informed representation of EEG signals, establishing a robust new paradigm for preictal–interictal EEG classification.

The present study is intended as a methodological evaluation of patient-specific representations within retrospectively constructed, fixed EEG networks. It does not evaluate zero-shot transfer to unseen patients or prospective seizure-warning performance. Cross-patient deployment would require an inductive representation that maps signals from a new patient into a common feature space learned exclusively from training patients.

\section{Results}\label{sec:results}

\subsection{Overview of the proposed framework}\label{subsec:overview}
\begin{figure}[!htbp]
\centering
\includegraphics[width=0.99\textwidth]{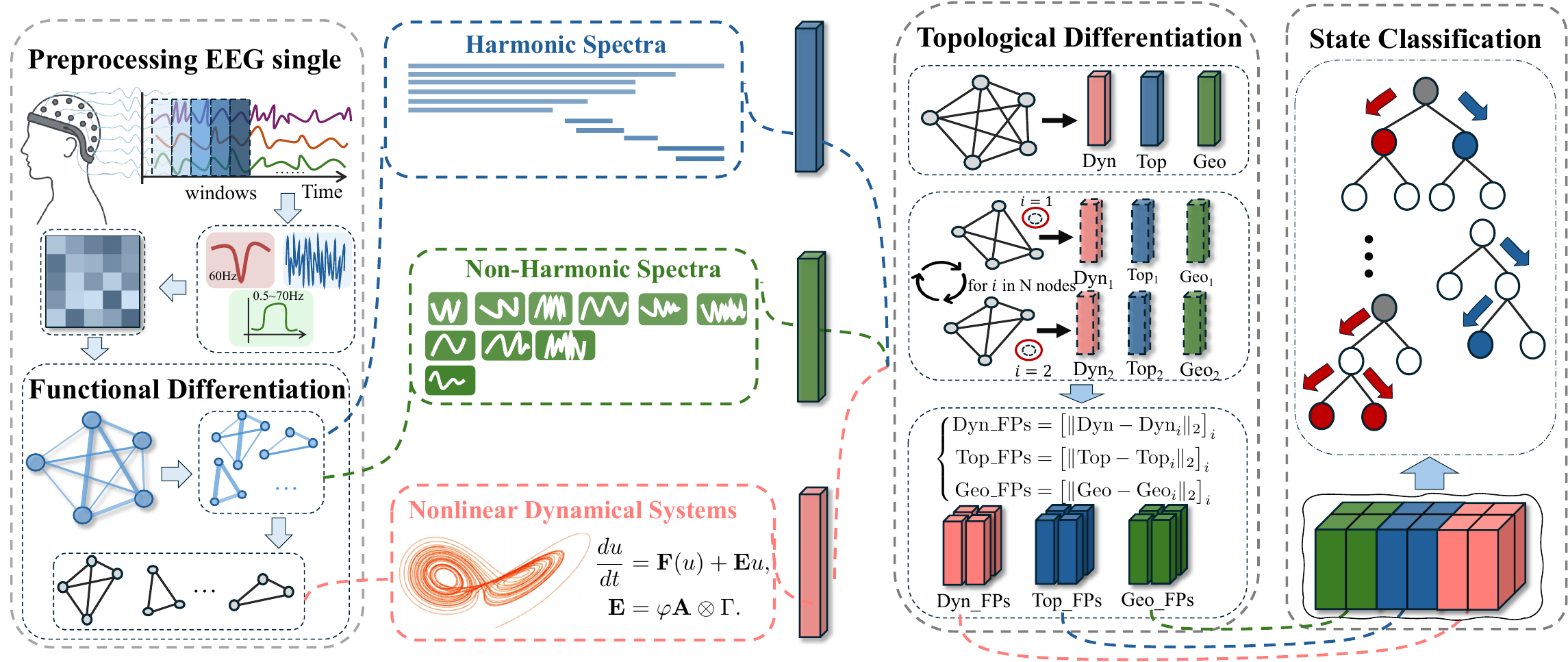}
\caption{Overview of the proposed CDRTL framework for patient-specific preictal-interictal EEG node classification. EEG signals are first preprocessed via temporal windowing, bandpass filtering (0.5–70 Hz), notch filtering (60 Hz), and normalization. Pairwise Pearson correlations are then computed to construct a weighted functional network. The network is partitioned into ten sub-networks based on decile intervals of edge weights, capturing different connectivity regimes. For each sub-network, nodes are modeled as nonlinear oscillators (e.g., Lorenz or R\"ossler), forming coupled dynamical systems. Persistent Laplacian–based multiscale analysis is applied to extract topological (Top) and geometric (Geo) spectral features, while statistical descriptors of dynamical trajectories provide dynamical features (Dyn). A topological differentiation scheme further derives node-wise features by measuring differences between original and node-removed global representations. Finally, fused node-level features from all sub-networks are used for downstream classification of preictal and interictal labeled channel-level nodes.}
\label{fig:framework}
\end{figure}

Figure~\ref{fig:framework} illustrates the CDRTL framework, which provides a general workflow for topology-enabled ML prediction regulated by system dynamics. First, the EEG data recorded from the patient's brain are transformed into a connectivity matrix representation after signal preprocessing. The matrix elements consist of the pairwise Pearson correlation coefficients between all EEG channels, which reflect the interaction strength or distance between nodes. The larger the correlation coefficient between the nodes, the stronger the interaction strength, and the shorter the distance between the nodes or brain regions. Next, a functional differentiation is implemented by partitioning correlation coefficients into ten decile intervals. Each interval corresponds to a sub-network composed of nodes whose interaction strength or distance falls within this interval. Consequently, these intervals yield ten sub-networks with comparable edge densities but different interaction strength. In each sub-network, each node represents an EEG channel. Class labels are not used in graph construction or feature extraction and are revealed only for the training nodes during supervised classifier fitting. The edges are defined by associations between samples, such as phase synchronization, coherence, or correlation. Unlike standard graph-based approaches, we treat each node in this network as a nonlinear oscillator. Specifically, we embed a nonlinear dynamical system in every node, such as Lorenz or R\"ossler dynamics, resulting in a coupled networked dynamical system. The interaction between nodes effectively modulates the dynamical behavior of the oscillators based on the brain's real-time functional state.

Second, a family of dynamical systems is constructed through a filtration process, with persistent Laplacian analysis applied to each sub-network. The coupled Lorenz equations are solved to obtain the time evolution trajectories of each oscillator. The statistical features extracted from $x$-axis trajectories represent dynamical fingerprints \DynFPs{}, summarizing the statistical responses of the coupled nonlinear oscillator network associated with interictal and preictal EEG states. Simultaneously, PL yields a dual set of fingerprints. Unlike conventional PH methods, PL supports the extraction of both harmonic spectra, corresponding to topological invariants, and non-harmonic spectra, corresponding to geometric and metric evolution characteristics, and they are summarized as topological fingerprints \TopFPs{} and geometric fingerprints \GeoFPs{}, respectively. The fusion of these three distinct fingerprint families \DynFPs{}, \TopFPs{}, and \GeoFPs{} constitutes the system's global features for each sub-network.

Third, the localized information for each node is derived by a topological differentiation method from the global topology-geometry-dynamics features. This process involves calculating the sensitivity of the global features to individual graph nodes. By doing so, we obtain node level local features that quantify the importance and role of each channel-level EEG node in the evolving preictal network. These localized features concatenated from ten sub-networks are then fed into a downstream ML prediction task to perform the binary preictal-interictal EEG node classification of the patient's state.   

The current framework performs patient-specific transductive node classification on a fixed network. The complete set of unlabeled channel-level signals is used to construct the correlation network and derive CDRTL features before cross-validation (CV). During each fold, held-out node labels are masked, and only the representations and labels of the training nodes are used to fit the classifier. The reported performance therefore characterizes classification of channel-level nodes within the fixed patient network. Classifying a newly arriving window or channel that did not participate in network construction would require an inductive out-of-network extension, which is outside the scope of the present work. Additionally, because each node level representation is defined relative to the complete patient-specific network, withholding an entire seizure would leave its windows without a training-derived feature mapping. A valid leave-one-seizure-out evaluation therefore also requires an inductive out-of-network extension that is not available in the current framework.

It is worth noting that the present study evaluates the ability of CDRTL-derived representations to distinguish retrospectively defined preictal and interictal EEG channels in an offline, patient specific setting. Although preictal state identification is an important issue of seizure prediction research, the current evaluation does not constitute validation of a complete prospective seizure-warning system. Event level sensitivity and false alarms per hour require seizure-wise independent testing on chronologically continuous recordings, together with predefined seizure prediction horizons, seizure occurrence periods, alarm-merging criteria, and refractory periods. Because the current dataset consists of selected and balanced temporal windows rather than continuous monitoring streams, these event level measures cannot be reliably inferred from the present channel-level results. Future work will extend the proposed representation framework to causal, seizure-wise continuous evaluation and examine its suitability for low-latency warning systems.

\subsection{Overall performance of patient-specific channel-level node classification}\label{subsec:accuracy}

To evaluate the channel-level node classification performance of the CDRTL framework, we used the CHB-MIT dataset, a widely adopted public benchmark for epileptic seizure prediction. It contains 23 patients, including 5 male patients aged 3 to 22 years and 18 female patients aged 1.5 to 19 years. Details of the description of dataset and preprocessing pipeline are provided in the Section \ref{subsec:dataset} of Methods and Table~S1 of the Supporting Information. After preprocessing, each patient contributed 26 nonoverlapping 10-s EEG windows, including 13 preictal and 13 interictal windows. Each window contained 18 EEG channels, and each channel specific time series was treated as one graph node, yielding 468 nodes per patient. These balanced pools provided the basis for network construction and downstream classification.

We embedded the EEG time-series data into a nonlinear dynamical framework by treating each graph node as a nonlinear oscillator, corresponding to one EEG channel. We then computed the Pearson correlation matrix over the patient-specific sample set and constructed a family of ten connectivity matrices for coupled oscillator sub-networks through the functional differentiation method. For each sub-network, a Lorenz system was introduced and the corresponding $x$-axis trajectories were collected. Summary statistics of these trajectories were summarized as dynamical fingerprints \DynFPs{}. In parallel, PL was applied to each sub-network to extract the topological invariants or topological fingerprints \TopFPs{} through the harmonic spectral analysis, and the geometric shape evolution or geometric fingerprints \GeoFPs{} through the non-harmonic spectral analysis. Concatenating these three fingerprint groups yielded the global features of the brain sub-network. Node-level difference vectors were then obtained using topological differentiation method and treated as local features of the sub-network, and the fused node-level fingerprints from ten sub-networks were used for EEG classification. Five classifiers, namely support vector machine (SVM), random forest (RF), $k$-nearest neighbors (KNN), logistic regression (LR), and gradient boosting decision tree (GBDT) with nested CV were employed to evaluate the capacity of the extracted features. The details of implementation of nested CV could be found in Section 4.2 of Methods. Classification performance was evaluated using accuracy, sensitivity, specificity, precision, F1-score, and the area under the receiver operating characteristic curve (AUC). The details of all hyperparameters used in various ML models can be found in Table S2 of the Supporting Information.

Using the complete nested CV procedure, CDRTL achieved a cohort-mean accuracy of 0.994 (95\% bootstrap confidence interval: 0.993, 0.995) and sensitivity of 0.995 (0.994, 0.997) across the 23 patients. The complete results for precision, specificity, F1-score, and AUC are provided in Table~S4 of the Supporting Information. Feature configuration, classifier type, and hyperparameters were selected exclusively within each inner three-fold loop, whereas the five outer folds were used only for evaluation. Additionally, across ten independent window-selection replicates, the cohort-mean accuracy was 0.994 $\pm$ 0.001 (mean $\pm$ standard deviation across replicates), with individual replicate means ranging from 0.993 to 0.995. This small variation indicates that the high cohort-level accuracy was reproducible across the tested window selections at the fixed budget of 13 windows per class. Table S22 of the Supporting Information reports the patient-level results for all ten replicates. These findings support stability within the evaluated sampling scheme and transductive setting.

\begin{figure}[tbp]
\centering
\vspace{-45pt}
\includegraphics[width=0.90\textwidth]{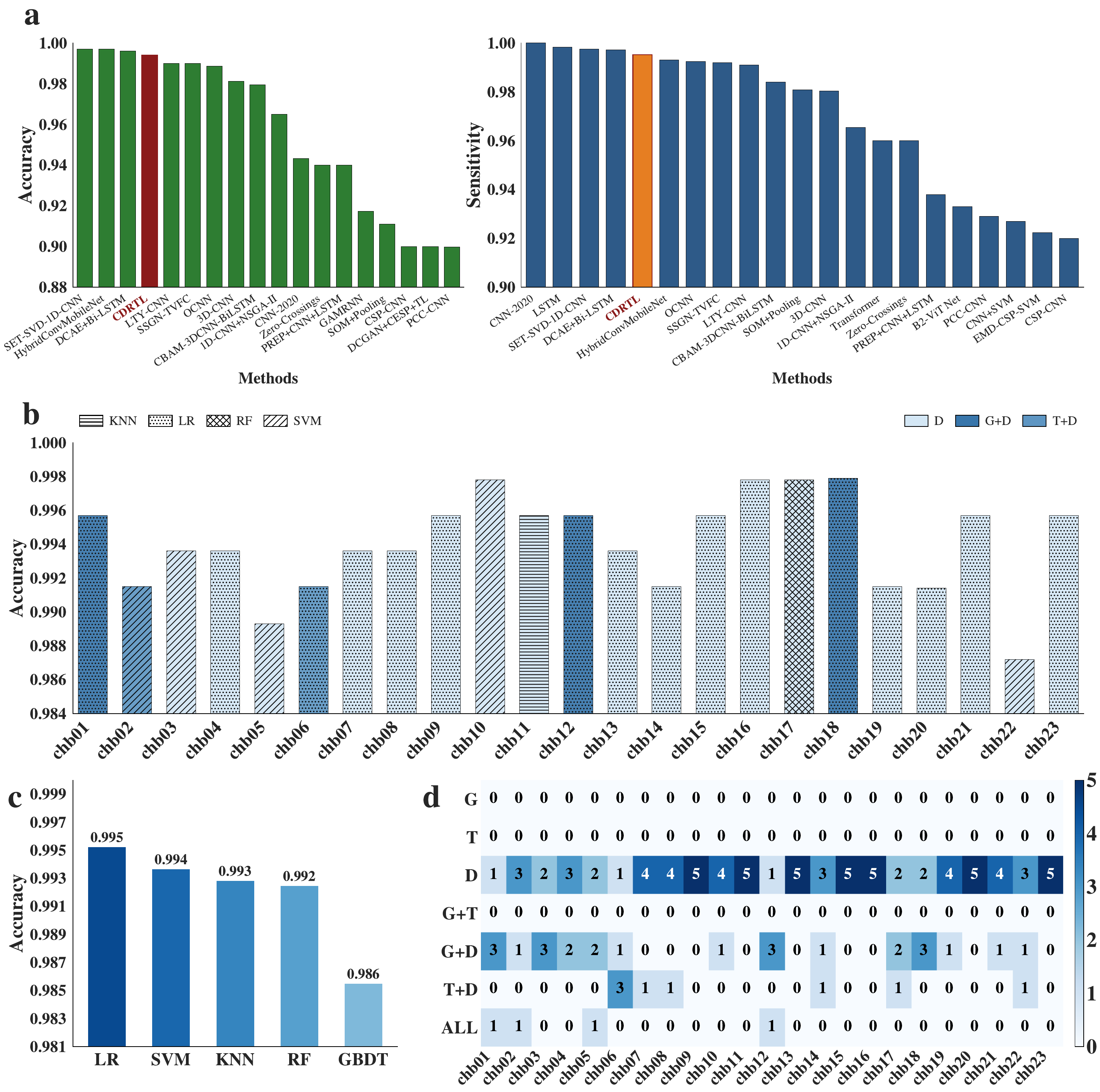}
\caption{Experimental performance across 23 patients on CHB-MIT dataset.
\textbf{a}, Contextual comparison of accuracy and sensitivity reported for representative methods on the CHB-MIT dataset. The CDRTL values, accuracy 0.994 and sensitivity 0.995, were obtained using the complete nested cross-validation (CV) procedure. Because the published studies used different sample definitions and validation protocols, the comparison is not intended as a direct performance ranking. 
\textbf{b}, Per-patient accuracy of the complete nested CV procedure, computed from the five outer test folds. Bar colors and hatching indicate the feature set and classifier most frequently selected by the inner loop across the five outer folds. When multiple configurations had the same frequency, the configuration with the higher corresponding mean inner-loop accuracy was displayed. If both frequency and accuracy were equal, one configuration was randomly selected. These annotations summarize selection frequency and do not imply that a single configuration generated all outer-fold predictions. The most frequently selected classifier include $k$-nearest neighbors (KNN), logistic regression (LR), random forest (RF), and support vector machine (SVM). Here, T, G, and D denote topological (\TopFPs{}), geometric (\GeoFPs{}), and dynamical (\DynFPs{}) fingerprints, respectively, ``+'' denotes feature concatenation, and chb01--chb23 denote the individual patient.
\textbf{c}, Mean patient-level accuracy across 23 patients for five fixed-classifier nested CV analyses, including LR, SVM, KNN, RF, and gradient boosted decision trees (GBDT). Within each outer-training partition, the classifier type was fixed, whereas the fingerprint combination and classifier-specific hyperparameters were selected using inner three-fold CV. Each patient-level value was calculated as the mean accuracy across the five outer test folds.
\textbf{d}, Heatmap of feature-configuration selection frequencies under nested CV with the classifier type fixed to LR. Each cell indicates the number of the five outer folds in which a given feature configuration was selected by the corresponding inner three-fold loop for each patient, with LR hyperparameters optimized within the same inner loop. The frequency ranges from 0 to 5, and each row sums to five. The color scale represents selection frequency rather than classification accuracy.}
\label{fig:result2}
\end{figure}

Figure~\ref{fig:result2}a places the primary nested CV results of CDRTL in the context of representative CHB-MIT studies. Its accuracy and sensitivity fall within the upper range of previously reported results. However, the included studies differ substantially in preprocessing, sample construction, preictal definitions, data-partitioning units, and validation protocols. The comparison should therefore be interpreted as a contextual benchmark rather than a protocol matched ranking of methods. The corresponding values and methodological information are summarized in Table~S3 of the Supporting Information.

To provide a more controlled reference, we additionally reimplemented SET-SVD-1D-CNN~\cite{ra_novel_2023} using the same data-processing and evaluation framework. Its mean accuracy was 0.534, whereas the primary nested-CV accuracy of CDRTL was 0.994, corresponding to an absolute difference of 0.460. Detailed results are provided in Tables~S20 and ~S21 of the Supporting Information.

Figure~\ref{fig:result2}b summarizes the per-patient accuracy of the complete nested CV procedure. The bar color and hatching denote the feature-classifier configuration most frequently selected by the inner loop. As the selected configuration could vary among outer folds, this annotation should be distinguished from the bar height, which represents the mean of the five resulting outer-test accuracies. Patient-level accuracy ranged from 0.987 to 0.998, with the highest values observed for chb10, chb16, chb17, and chb18. All displayed modal feature configurations contained \DynFPs{}. \DynFPs{} alone was selected for 18 patients, whereas G+D and T+D were selected for three and two patients, respectively. This consistent preference suggests that \DynFPs{} provide the principal discriminative information within the evaluated feature families.

Classifier selection also varied across patients. LR was the modal classifier for 16 patients, followed by SVM for five patients, KNN for one patient, and RF for one patient. The frequent selection of LR may indicate that the extracted representations often support an approximately linear separation of preictal and interictal channel-level nodes under the present within-patient transductive setting. In contrast, the selection of SVM, KNN, or RF for several patients may reflect patient-dependent differences in feature-space geometry. Overall, these findings support the value of adaptive configuration selection, although they do not imply that the displayed modal configuration generated all outer-fold predictions or that it generalizes to unseen patients. Table~S6 of the Supporting Information reports the corresponding metrics and selected configurations in full.

To examine whether the discriminative performance depended strongly on the classifier backend, Figure~\ref{fig:result2}c compares five fixed-classifier nested CV analyses. For each classifier and outer fold, the classifier type was held fixed, whereas the fingerprint combination and classifier-specific hyperparameters were selected exclusively by the inner three-fold loop. The selected configuration was then refitted on the outer-training partition and evaluated on the corresponding outer-test partition. Patient-level accuracy was calculated as the mean across the five outer folds and subsequently averaged across the 23 patients.

LR yielded the highest observed mean accuracy of 0.995 (0.994, 0.996), followed by SVM at 0.994 (0.992, 0.995), KNN at 0.993 (0.991, 0.994), RF at 0.992 (0.991, 0.994), and GBDT at 0.986 (0.983, 0.988). Because the confidence intervals of LR, SVM, KNN, and RF overlap, their small numerical differences should not be interpreted as evidence of a statistically significant ranking. Nevertheless, the consistently high accuracies across these classifier families indicate that the extracted fingerprints retain strong discriminative information without depending on a particular classification algorithm. Summary statistics and detailed patient-level results are provided in Tables~S5 and ~S7 of the Supporting Information, respectively.

Figure~\ref{fig:result2}d illustrates the stability of feature-configuration selection under nested CV when the classifier type was fixed to LR. For each patient, each cell represents the number of the five outer folds in which the corresponding feature configuration was selected by the inner loop. Consequently, the frequencies in each row sum to five. D alone showed the highest selection frequency for 18 of the 23 patients, whereas G+D was most frequently selected for chb01, chb03, chb12, and chb18, and T+D for chb06. Thus, the modal configuration contained D for every patient. For some patients, selections were distributed across multiple configurations, possibly because these configurations yielded similar inner-validation performance or because selection was sensitive to changes in the outer-training subset. Importantly, the heatmap quantifies selection consistency rather than outer-test accuracy or universal feature superiority. The predominance of D supports their central discriminative contribution under the fixed-LR setting.

As a complementary fixed-configuration analysis, \DynFPs{} and LR were held constant and evaluated using conventional stratified five-fold CV, without selection among feature configurations or classifier types. Accuracy, precision, sensitivity, specificity, and F1-score were each 0.996 ± 0.002 (0.994, 1.000). The close agreement with the primary nested CV result indicates that high node-level discrimination can also be obtained with this single fixed configuration. This analysis is complementary and does not replace the nested CV estimate as the primary performance result.

\section{Discussion}\label{sec:discussion}
The classification performance should be interpreted within the patient-specific transductive setting. Held-out nodes contribute their unlabeled EEG signals to network construction and representation extraction before CV; only their labels are withheld from classifier training and selection. Nested CV separates model selection from outer-fold evaluation, but does not make representation construction independent of the held-out nodes. The resulting estimates therefore concern label discrimination within an already observed network, rather than generalization to new EEG windows, seizure events, or patients.

\subsection{Visual assessment of feature-space discriminability and class separation}\label{subsec:dis_manifold}

\begin{figure}[!htbp]
\centering
\includegraphics[width=0.70\textwidth]{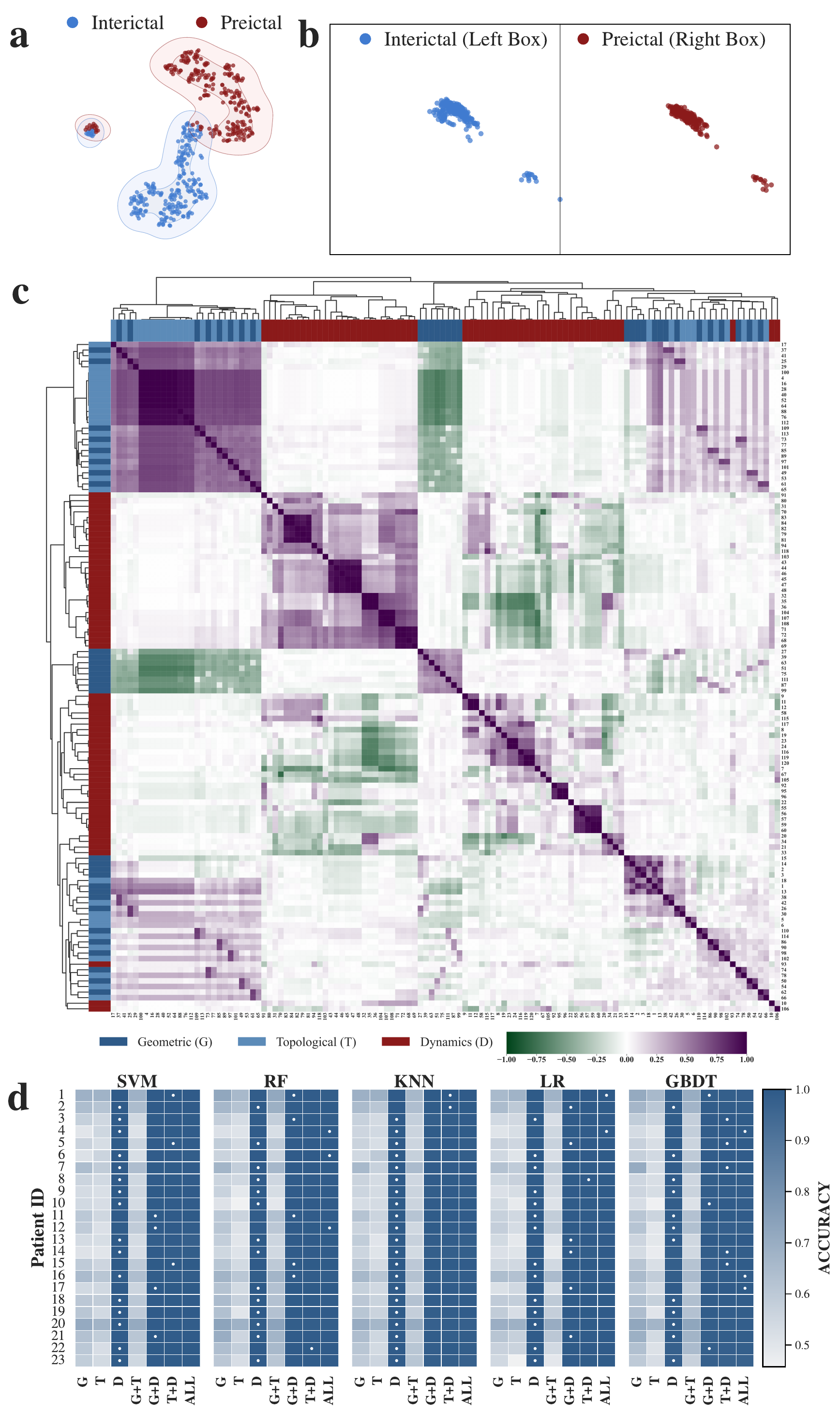}
\caption{Feature-space visualization, correlation structure, and exploratory feature-ablation analysis.
\textbf{a}, t-SNE visualization of the proposed feature representation and three contour levels illustrate the low-dimensional class-boundary structure.
\textbf{b}, RS (residue-similarity) score map, summarizing feature-manifold separability and compactness.
\textbf{c}, Hierarchically clustered correlation matrix of the 120-dimensional  feature representation, comprising 30 topological dimensions (T), 30 geometric dimensions (G), and 60 dynamical dimensions (D). 
\textbf{d}, Exploratory feature-ablation heatmaps showing stratified five-fold cross- validation accuracy for each prespecified classifier-feature pair across the 23 patients. Identical data partitions were used for all pairs. This analysis compares fixed configurations descriptively and is not used as an independent generalization estimate after feature or classifier selection. The analysis in panels \textbf{a}, \textbf{b}, and \textbf{c} are for patient chb01.}
\label{fig:result3}
\end{figure}

To qualitatively assess the discriminative capability of the feature set (\TopFPs{}, \GeoFPs{} and \DynFPs{}), a t-SNE visualization was performed for subject chb01 in Figure~\ref{fig:result3}a. The resulting map shows a striking spatial separation between preictal and interictal manifolds with minimal overlap. This high degree of inter-class separability provides visual evidence for the exceptional classification performance observed across the patients, confirming that the proposed fingerprints effectively capture the distinct neural dynamics preceding a seizure. The complete manifold gallery for 23 patients is provided in Section~S6 and Figure~S1 of the Supporting Information.

To further evaluate the classification reliability, taking subject chb01 as an example, an R-S (Residue-Similarity) score analysis was conducted in Figure~\ref{fig:result3}b, which is a novel visualization tool used to evaluate the clustering and classification performance of the proposed framework. $X$-axis means residue score, $R$, representing the interclass distance, and $y$-axis means similarity score, $S$, indicating the intraclass similarity. In both the Interictal and Preictal boxes, the majority of the data points are concentrated in the upper regions of the map. This indicates that most samples for patient chb01 possess high similarity and residue scores, which directly correlates with the high classification accuracy (99.6\%) observed in the quantitative results. The details of R-S score of the remaining patients are shown in Table~S8 and Figure~S2 of the Supporting Information.

\subsection{Feature complementarity, ablation, and inter-patient heterogeneity}\label{subsec:dis_orthogonal}

To investigate the internal structure and redundancy of the fused feature space, the hierarchical clustered correlation heatmap of the 120-dimensional feature space for patient chb01 is shown in Figure~\ref{fig:result3}c. It reveals a clear structural organization among topological (T), geometric (G), and dynamical (D) features, as evidenced by both the block-wise correlation patterns and the dendrogram arrangement. The proposed feature representation is a composite vector constructed through the fusion of 30-dimensional topological fingerprints (light blue color), 30-dimensional geometric fingerprints (dark blue color), and 60-dimensional dynamical fingerprints (red color). The numbers shown in bottom- and right-axis are the labels of feature. The scale of colorbar ranges from dark green ($-1.00$, strong negative correlation) to white ($0.00$, no correlation) and dark purple ($1.00$, strong positive correlation). Specifically, the topological features exhibit a markedly higher level of internal consistency, with a mean within-group Pearson correlation of 0.437, indicating the redundancy and suggesting that these features capture closely related structural properties of the underlying EEG-derived topology. In contrast, the geometric and dynamical feature sets display much weaker internal correlations, with mean values of 0.081 and 0.059, respectively, implying that these features are more diverse and less redundant within their own groups. The details of the complete numerical entries underlying Figure~\ref{fig:result3}c are provided in supplementary tables in the GitHub repository.

The cross-group correlation analysis further highlights the relative independence among feature categories. The correlations between dynamical features and the other two groups are nearly zero (-0.004 for G-D and -0.003 for T-D), indicating strong orthogonality and suggesting that dynamical features encode information that is largely complementary to both geometric and topological fingerprints. Meanwhile, the slightly positive correlation between geometric and topological features (0.088) suggests a weak but non-negligible association, potentially reflecting partial overlap in the structural information they capture.

These correlation patterns are consistent with the hierarchical clustering structure observed along the axes, where topological features tend to form tightly grouped clusters, while geometric features exhibit looser clustering and dynamical features appear more dispersed. This organization indicates that the topological feature space is highly structured but potentially redundant, whereas the geometric and especially the dynamical feature spaces provide more heterogeneous and complementary representations.

In general, the observed correlation structure demonstrates a favorable balance between redundancy and complementarity between feature types. Although topological features contribute to a concentrated and internally consistent representation, geometric and dynamical features introduce additional, largely independent information. These correlation patterns describe the relationships among the feature families, but do not establish that combining them improves classification performance over \DynFPs{} alone.

Figure~\ref{fig:result3}d provides an exploratory feature-ablation analysis rather than a primary estimate of post-selection generalization. Each of the 35 prespecified classifier-feature pairs was evaluated using identical CV partitions, allowing controlled descriptive comparisons across feature families. This performance landscape is evaluated in five distinct ML classifiers, such as SVM, RF, KNN, LR, and GBDT. The color intensity gradient, transitioning from light blue to dark blue, maps to the classification accuracy, while the white dots denote, for each patient and fixed classifier, the feature configuration with the highest CV accuracy among the seven configurations evaluated. These markers summarize a retrospective within-classifier comparison rather than feature sets selected through an independent nested-validation procedure. The details of optimal accuracy for each patient with five classifiers and the full numerical results for Figure~\ref{fig:result3}d are tabulated in supplementary tables of the GitHub repository. 

As summarized in Table S9 of the Supporting Information, configurations without D (G, T, and G+T) yielded cohort-mean accuracies ranging from 0.541 to 0.616, whereas configurations containing D achieved 0.988–0.996 across the five classifiers. This dramatic contrast identifies the Lorenz-derived nonlinear dynamical features (D) as the strongest discriminative component under the present transductive node classification protocol. 

The distribution of the optimal performance markers (white dots) further elucidates the interplay between feature selection and patient heterogeneity. Most of these optimal markers are concentrated strictly within the standalone "D" column, particularly for classifier KNN. Specifically, the ratios of patients obtaining optimal accuracy with dynamical features alone among five classifiers are 16/23, 14/23, 21/23, 14/23, and 12/23, respectively. The exploratory comparisons in Figure~\ref{fig:result3}d indicate that \DynFPs{} alone often achieved the highest observed accuracy. The nested-CV selection summaries in Figures~\ref{fig:result2}b and \ref{fig:result2}d likewise show a predominance of \DynFPs{}, although fused configurations were selected for some patients. These selections suggest patient-dependent utility of the additional features, but their frequency does not quantify the incremental accuracy gain or establish a reliable advantage over \DynFPs{} alone.

\TopFPs{} and \GeoFPs{} are retained as explicit descriptors of network topology and geometry, derived from the harmonic and non-harmonic persistent-Laplacian spectra, respectively. They allow structural information to be examined alongside the network responses summarized by \DynFPs{}. Although \DynFPs{} comprise simple summary statistics, these statistics are computed from coupled oscillator trajectories and their changes under node removal, rather than directly from raw EEG signals. The integrated representation therefore provides a framework for comparing structural and dynamical information, without requiring all three feature families to contribute equally or to be included in every selected classifier. The present results support \DynFPs{} as the principal predictive component and the spectral fingerprints as candidate supplementary descriptors.

Retaining the spectral fingerprints nevertheless requires additional persistent-Laplacian computations across scales and node-removal perturbations. We have not benchmarked their incremental runtime or memory requirements, and the observed selection patterns are insufficient to establish that their predictive gains justify this computational cost. A \DynFPs{}-only implementation could omit these spectral computations while retaining network construction and dynamical simulation. A paired comparison of \DynFPs{}-only and fused configurations under identical nested-CV partitions, together with runtime and memory measurements, is needed to determine when the additional descriptors are worthwhile.

\subsection{Control analyses of dynamical fingerprints, connectivity, and coupling}
To determine whether the performance of \DynFPs{} could be reproduced by a generic nonlinear transformation, we replaced the Lorenz transformation with sigmoid and tanh controls while retaining the same patient cohort, graph filtration, feature dimensionality and summary statistics, node-removal differentiation, CV partitions, classifier protocol, and bootstrap procedure. The sigmoid control produced near-chance performance, with a mean accuracy of 0.508 (0.497, 0.527) and a mean AUC of 0.477 (0.395, 0.559). The tanh control retained moderate discriminative capacity, achieving a mean accuracy of 0.774 (0.717, 0.828) and a mean AUC of 0.791 (0.716, 0.857), but remained markedly below the Lorenz-based representation, which achieved an accuracy of 0.994 (0.993, 0.995) and an AUC of 1.000 (0.999, 1.000). The details of implementation of these two controls and evaluation results are provided in Tables~S14 and ~S15 of the Supporting Information.

In order to demonstrate the robustness of the proposed framework CDRTL with respect to the choice of chaotic dynamics, we consider two dynamical systems, i.e. Lorenz and R\"{o}ssler systems. Specifically, the two systems correspond to mean predictive accuracies of 0.994 (0.993, 0.995) and 0.998 (0.997, 0.999) in 23 patients, respectively, under the same nested CV protocol, which verifies that the robustness of CDRTL is independent of specific dynamical systems. The details of the robustness analysis of dynamical system choice are given in Section~S5 and Tables S10, S11, S12 of the Supporting Information. These two controls address complementary questions that sigmoid and tanh assess whether an arbitrary simple nonlinearity is sufficient, whereas Rössler assesses dependence on the particular chaotic oscillator. Together, the results support the value of coupled nonlinear temporal evolution and demonstrate robustness to oscillator choice.

To further examine whether the strong performance of \DynFPs{} was primarily determined by the organization of the empirical connectivity matrix, we performed a connectivity-weight shuffling control for all 23 patients. For each patient, ten independent surrogate matrices were generated, and the shuffled performance was averaged across these repetitions. Specifically, the strict upper-triangular off-diagonal elements of the Pearson matrix were randomly permuted and mirrored to preserve symmetry, while all subsequent procedures remained unchanged. Using fixed \DynFPs{} and classifier LR, the original and shuffled matrices yielded mean accuracies of 0.982 (0.976, 0.987) and 0.980 (0.974, 0.986), respectively. This comparable result suggests that \DynFPs{} are relatively insensitive to the exact node-pair arrangement of connectivity weights and may emphasize collective network responses or distribution-level connectivity information. Nevertheless, because the weight distribution and the Lorenz-based nonlinear transformation were retained after shuffling, this experiment does not independently establish a chaos-specific mechanism or rule out contributions from generic nonlinear transformations.

Additionally, we examined the contribution of inter-node coupling in chb01 for simplicity. Within the shuffled-connectivity setting, eliminating coupling by setting ($\phi=0$) reduced accuracy from 0.983 (0.979, 0.987) under the coupled condition to 0.516 (0.500, 0.531), approaching chance level. The preserved performance after weight shuffling, together with its collapse after coupling removal, suggests that \DynFPs{} primarily characterize coupling-dependent collective temporal responses rather than the precise empirical placement of individual connectivity weights. Nevertheless, this control demonstrates the importance of coupling between channel-level nodes. The detailed shuffling control results could be found in Table~S16 of the Supporting Information.

\subsection{Effects of multi-scale network interaction and cumulative feature fusion }\label{subsec:dis_coupling}

\begin{figure}[!htbp]
\centering
\includegraphics[width=0.99\textwidth]{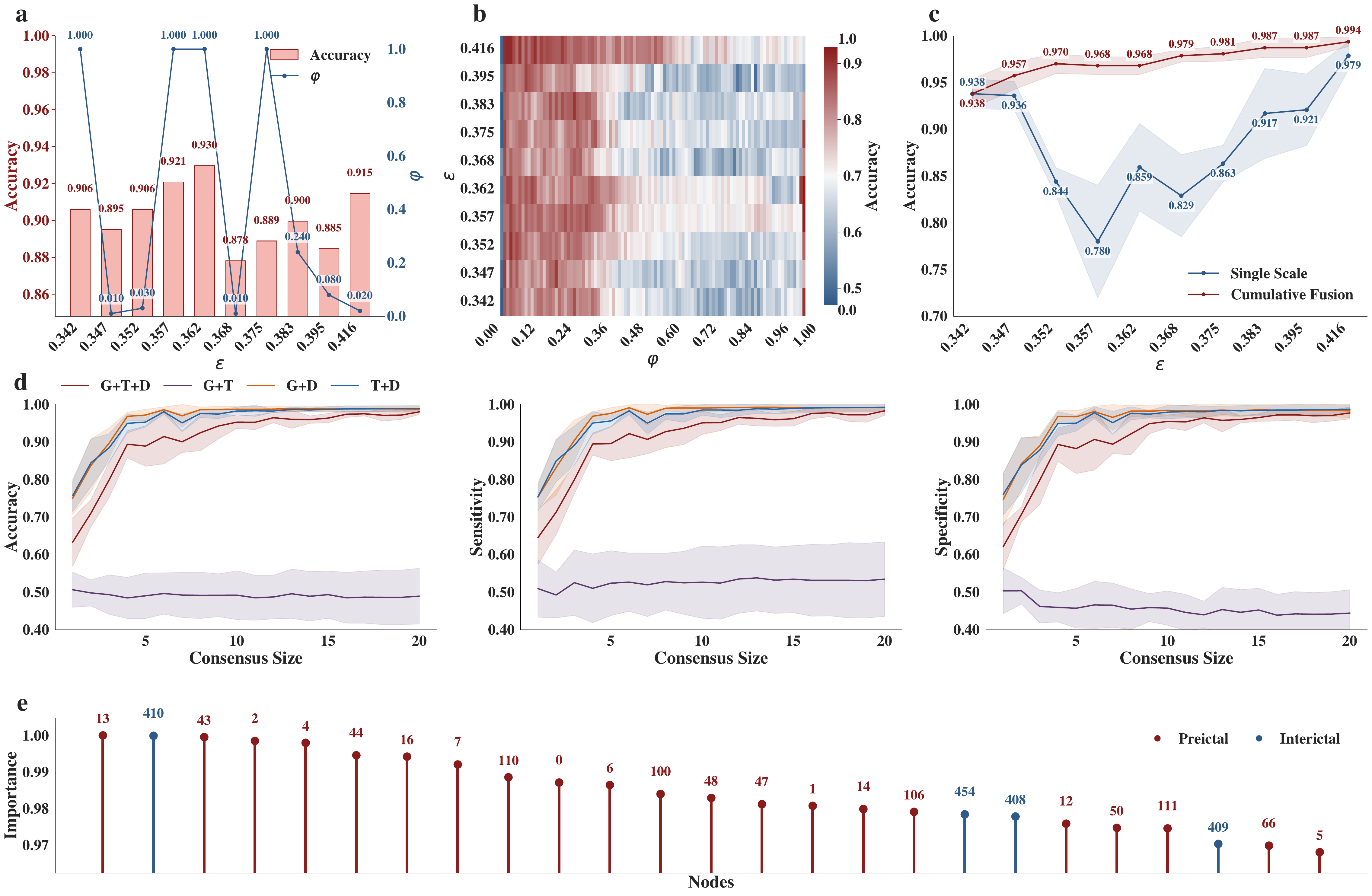}
\caption{Multiscale network analysis, ensemble stability, and patient-level structural interpretation.
\textbf{a}, Dual-$y$-axis analysis of model performance and network global coupling strength across different nodal distance thresholds.
\textbf{b}, The joint influence of global coupling strength $\varphi$ and distance threshold $\varepsilon$ on classification accuracy. 
\textbf{c}, The impact of multi-scale topological feature integration on classification accuracy with given network coupling strength 0.42. The chosen classifier in machine learning model is RF and the feature set is ALL, that is the concatenation of \TopFPs{}, \GeoFPs{}, and \DynFPs{}.
\textbf{d}, Consensus size analysis under a fixed LR model, showing how accuracy, sensitivity, and specificity vary as consensus size increases, where consensus size denotes the number of aggregated LR models.
\textbf{e}, Top-25 most important node ranking in the sub-network with largest decile threshold 0.416.
The analysis in panels \textbf{a}, \textbf{b}, \textbf{c}, \textbf{d} and \textbf{e} are for patient chb01.
}
\label{fig:result4}
\end{figure}

Figure~\ref{fig:result4}a presents a dual-$y$-axis analysis of model performance and network coupling strength across edge weights or nodal distance thresholds for subject chb01. The red bars (left $y$-axis) denote classification accuracy, while the blue dotted curve (right $y$-axis) represents the global coupling strength $\varphi$ between network nodes. The $x$-axis corresponds to the lower bounds $\varepsilon$ of ten decile intervals of nodal distance, where each value corresponds a distinct sub-network composed of nodes whose distance falls within the corresponding interval. These intervals derive ten sub-networks with comparable edge densities but different nodal distance. The nodal distance can also be seen as the interaction strength among nodes which decays with the distance between two nodes. In contrast, the right $y$-axis represents the coupling strength between nodes, which characterizes the dynamical interaction intensity and is physically different from the distance-based measure used for subgraph generation. The classifier used in the ML model is RF and the feature selection is ALL (\TopFPs{}+\GeoFPs{}+\DynFPs{}) in the prediction. 

Figure~\ref{fig:result4}a indicates a highly nonlinear dependency between the chosen distance thresholds $\varepsilon$, the resulting network dynamics, and classification performance. Notably, the model achieves its maximum classification accuracy 0.930 at a distance threshold of 0.362, which precisely coincides with a maximal coupling strength of $\varepsilon = 1.0$. A similarly robust performance is observed at the adjacent distance threshold of 0.357, where the accuracy remains high and the coupling strength is sustained at its maximum.

However, Figure~\ref{fig:result4}a also indicates that maximal coupling strength alone does not universally guarantee superior classification. For instance, at the distance threshold of 0.375, with the maximum coupling strength 1.0, yet the accuracy experiences a decrease to 0.889. Conversely, the sharpest decrease in accuracy, occurring at the distance threshold 0.368 with minimum accuracy 0.878, aligns with a near-zero coupling strength. This profound drop suggests that at this specific nodal distance level, the lack of sufficient network interaction critically degrades the discriminative power of the extracted topological invariants. Ultimately, these findings validate the efficacy of the interaction specific PL method. By navigating the balance between the degree of node correlation and topological information, the framework successfully isolates optimal structural states, such as the configuration at the decile threshold of 0.362 where the underlying neural coupling dynamics differentiates more effectively between preictal and interictal electrophysiological patterns.

Figure~\ref{fig:result4}b illustrates the joint influence of global coupling strength $\varphi$ and distance thresholds $\varepsilon$ on classification accuracy with the classifier RF and the feature set ALL (\TopFPs{}+\GeoFPs{}+\DynFPs{}). The $x$-axis represents the coupling strength between network nodes, while the $y$-axis corresponds to the lower bounds of ten decile intervals of nodal distance. The color gradient, transitioning from dark blue (lower accuracy) to deep red (peak accuracy approaching 1.0), reveals a highly non-uniform and parameter-sensitive performance landscape.

A primary observation is that robust classification accuracy is predominantly concentrated within the regime of relatively weak coupling strengths ($\varphi \le 0.42$) across the vast majority of decile intervals. This trend suggests that excessive dynamical interaction between nodes may homogenize the underlying neural signals, washing out the discriminative electrophysiological features required to separate preictal from interictal states. Conversely, weaker coupling preserves these critical, state-specific dynamics. Furthermore, the classification capability is heavily modulated by the chosen subgraph architecture. Structural configurations corresponding to intermediate distance bounds, most notably at $\varepsilon = 0.357$ and $\varepsilon = 0.362$, demonstrate significant enhanced robustness. These specific topologies sustain elevated accuracy levels (deep red) over a considerably broader spectrum of coupling strengths compared to upper (e.g., $\varepsilon=0.395$) or lower (e.g., $\varepsilon=0.347$) distance bounds, which rapidly degrade into the low-accuracy blue regions as $\varphi$ increases.

The results suggest that neither strong correlation nor strong coupling alone is sufficient, instead, an appropriate balance between the two is required to maintain both structural diversity and meaningful interaction patterns. This finding supports integrating correlation-based network construction with dynamical modeling, as it enables the extraction of complementary topological and dynamical information that enhances the separability of preictal and interictal states. The full numerical grid underlying Figure~\ref{fig:result4}b is provided in supplementary tables in the GitHub repository.

To further evaluate the impact of multi-scale topological feature integration, Figure~\ref{fig:result4}c illustrates the classification accuracy under a fixed network coupling strength of 0.42 with the classifier RF and the feature set ALL (\TopFPs{}+\GeoFPs{}+\DynFPs{}). The $x$-axis corresponds to the lower bounds of ten decile intervals of node interaction strength, where each interval defines a sub-network composed of edges whose weight values fall within the specified range. The blue markers represent the accuracy achieved using features extracted from each individual sub-network, while the red markers indicate the performance obtained by cumulatively fusing features from multiple intervals up to the given threshold.

Figure~\ref{fig:result4}c shows that single-scale performance marked by blue color varies substantially from 0.780 to 0.979 across different interaction intervals, indicating that subgraphs defined at different interaction levels capture heterogeneous and complementary information. This variability reflects the fact that restricting the analysis to a narrow interaction range limits the representation of the underlying system dynamics and structural diversity. In contrast, the cumulative fusion strategy marked by red color demonstrates a clear and consistent improvement in accuracy as more intervals are incorporated. Specifically, starting from an initial accuracy of 0.938, the performance increases steadily and reaches 0.994 when all ten intervals are combined. This monotonic trend indicates that features from different interaction regimes provide complementary information, and their integration effectively enhances the discriminative capacity of the model. The improvement also suggests that the interaction-strength-based decomposition captures multi-scale characteristics of the dynamical system, and aggregating these multi-scale features enables a more comprehensive representation of the underlying EEG dynamics.

\subsection{Effect of consensus size on ensemble classification stability and performance}\label{subsec:dis_consensus}

To rigorously evaluate the predictive stability and efficacy of different feature combinations, an ensemble-based consensus modeling approach with a fixed classifier LR on patient chb01 is employed, as depicted in Figure~\ref{fig:result4}d with three metrics, including accuracy, sensitivity, and specificity. For each feature domain, geometric (G), topological (T), and dynamical (D) fingerprints, a model pool of 20 candidate models is generated through bootstrap resampling. The consensus size on the $x$-axis dictates the number of models randomly drawn from each model pool to form the ensemble. For instance, a consensus size of $n$ for the feature sets combination (T+D) entails averaging the predictions of $n$ topological models and $n$ dynamical models. To mitigate random sampling bias and isolate the true underlying performance, this random drawing process is iterated 10 times and five-fold CV are performed for each consensus size, yielding the mean performance metrics alongside their corresponding variance bounds shown by the shaded background. The details of Figure~\ref{fig:result4}d can be found in Table~S13 of the Supporting Information.

A critical analysis of the resulting trajectories reveals the indispensable nature of the dynamical (D) features in the classification task. The G+T fusion, which relies solely on structural features and lacks dynamical information, completely fails to discriminate between the brain states. Its performance stagnates near a random-chance baseline of 0.5 across all three metrics, and its variance remains high, regardless of the consensus size.
In contrast, any consensus combination incorporating the dynamical features (G+D, T+D, or G+T+D) exhibits a rapid improvement in predictive power. At small consensus sizes (e.g., 1 to 5), the performance of these D-inclusive ensembles is moderately lower with larger standard deviations, reflecting the inherent variance and potential overfitting of individual base models. However, as the consensus size increases from 5 to 10, three metrics sharply increases and are eventually close to near-optimal level. This asymptotic convergence demonstrates that a relatively small ensemble of approximately 5 to 10 models per feature domain is sufficient to smooth out individual model biases and achieve a highly stable, robust prediction.

\subsection{Topological-differentiation-based node-importance analysis}\label{subsec:dis_vitality}

In the present work, the definition of node importance in the topological differentiation analysis is that, after we obtained three fingerprint groups, \DynFPs{}, \GeoFPs{}, and \TopFPs{} in one sub-network, we then vectorized these features, generating a comprehensive representation of EEG network's characteristic. We calculate the Euclidean distance between vectorized features before and after the deletion of a node. This distance serves as a metric for assessing the structural and functional importance or impact of the node's absence within the network. The illustration of topological differentiation analysis is given in Figure~S3 of the Supporting Information.

The topological differentiation analysis suggests that preictal nodes are not only distinct in feature values but may also exhibit greater structural influence on the organization of the brain network. As illustrated by the top-25 stem plot in Figure~\ref{fig:result4}e for patient chb01 at largest decile threshold of 0.416, the higher importance ranks are more densely occupied by preictal nodes than by interictal nodes in the sub-network where interactions are strongest between nodes. The overrepresentation of preictal nodes (21/25) suggests that the preictal state is associated with a subset of samples whose removal induces disproportionately large changes in the topological, geometric, and dynamical fingerprints. This shifts the interpretation beyond simple class separability toward a form of state-dependent network vulnerability. The numbers on the top of black dots in Figure~\ref{fig:result4}e are sample IDs. The node importance ranking for the remaining patients can be found in Figure S4 of the Supporting Information.

From a mechanistic perspective, preictal activity may correspond to a partially reorganized state in which a limited set of structurally influential samples anchors the evolving graph. Within a nonlinear dynamical framework, this observation is consistent with the hypothesis that the system moves from a relatively diffuse background state toward a more coordinated pre-seizure regime. In that setting, topological differentiation applied to specific preictal nodes would be expected to yield larger system-wide changes, reflecting their greater contribution to the emerging collective organization.

This interpretation is further supported by the subject-level enrichment analysis, which shows that preictal nodes account for $63.9\% \pm 8.8\%$ (mean ± s.e.m.) of the top 20 most sensitive ranks across 23 patients. Detailed patient-level counts and the corresponding statistical procedure are provided in Table~S17 of the Supporting Information. This result improves interpretability by suggesting that the framework preferentially identifies graph samples associated with the preictal state rather than functioning solely as a black-box binary classifier. Although validation in larger and more diverse datasets is still required, the finding points to a plausible connection between predictive modeling and the mechanistic characterization of seizure evolution.

\subsection{Robustness to simulated EEG noise and physiological artifacts}\label{subsec:noise_robustness}

\begin{figure}[!tbp]
\centering
\adjustbox{width=0.99\textwidth}{%
\includegraphics{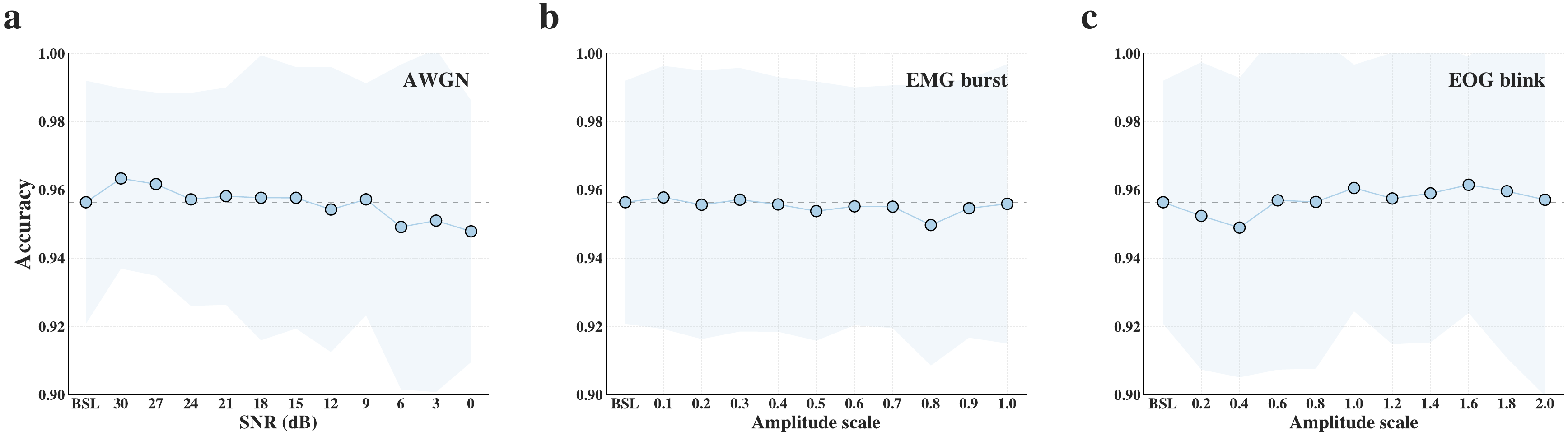}%
}
\caption{Robustness to simulated EEG noise and physiological artifacts.
\textbf{a}, Accuracy under additive white Gaussian noise (AWGN) with SNR decreasing from 30 to 0 dB.
\textbf{b}, Accuracy under electromyographic (EMG) burst artifacts with the amplitude scale increasing from 0.1 to 1.0.
\textbf{c}, Accuracy under electrooculographic (EOG) blink-like artifacts with the amplitude scale increasing from 0.2 to 2.0.
For \textbf{a}--\textbf{c}, noise was added after window-wise standardization, followed by feature re-extraction and LR classification using G+D+T (ALL); BSL denotes the clean-data baseline accuracy averaged over 23 patients under five-fold cross-validation. The standard deviation is presented by shaded background.
}
\label{fig:noise}
\end{figure}

To further evaluate the reliability of the proposed fingerprints under acquisition noise and physiological artifacts, we conducted a noise robustness analysis after the window-wise standardization step. This method fixed the classifier as LR and used the G+D+T (ALL) fingerprint combination with five-fold CV for all clean and noisy conditions shown as baseline point (BSL) in Figure~\ref{fig:noise}a--c, so that the performance changes could be attributed to the perturbation level rather than model selection. Specifically, three controlled perturbations were carried out to all standardized EEG channels before feature extraction, and the corresponding fingerprints were then recomputed for classification evaluation. First, additive white Gaussian noise (AWGN) was introduced with the target signal-to-noise ratio (SNR) varying from 30 to 0 dB, where lower SNR indicates stronger noise and SNR = 0 dB corresponds to noise power comparable to signal power. This design follows prior EEG studies that used additive Gaussian perturbations for robustness, distribution-shift, or synthetic-noise analysis \cite{e19080385,NEURIPS2022_8511d06d,10.3389/fninf.2025.1521805}. Second, electromyographic (EMG) artifacts were simulated as transient band-limited bursts with local channel involvement, motivated by the spectral and spatial characteristics of myogenic contamination \cite{pope2022managing}. Third, electrooculographic (EOG) artifacts were modeled as blink-like transient pulses with larger anterior-channel weights, consistent with the frontal dominance of ocular artifacts in EEG recordings \cite{bioengineering11101018}. The details of implementations of noise addition can be found in Section~S8 of the Supporting Information.

As shown in Figure~\ref{fig:noise}a, the AWGN experiment shows the clearest noise-dependent downward tendency. At high SNR levels, the classification accuracy remains comparable to the clean baseline of 0.956, reaching 0.963 and 0.962 at 30 and 27 dB, respectively. As SNR decreases, corresponding to stronger noise, the accuracy gradually declines with small fluctuations, and the lowest value of 0.948 is observed at 0 dB, which decreases 0.9\% of baseline. It suggests that CDRTL model exhibits good overall stability and reliability under the influence of noise in the data.

The EMG burst experiment in Figure~\ref{fig:noise}b shows smaller and less monotonic fluctuations. Across amplitude scales from 0.1 to 1.0, the mean accuracy varies within a narrow range from 0.950 to 0.958. The lowest value appears at the amplitude scale of 0.8, whereas the accuracy returns to 0.956 at the largest amplitude scale. This pattern suggests that transient myogenic bursts produce only limited degradation under the present perturbation setting, and the chosen fingerprint representation remains close to its clean-data baseline across most EMG levels.

For EOG blink-like artifacts, Figure~\ref{fig:noise}c also does not show a monotonic accuracy decrease with increasing amplitude. The accuracy initially drops to 0.949 at amplitude 0.4, but then returns to the baseline level and reaches values above 0.960 at amplitudes 1.0 and 1.6. Even at the strongest EOG amplitude of 2.0, the accuracy remains 0.957, which is almost equivalent to the clean baseline 0.956. These results indicate that the classification performance is relatively stable under blink-like transient artifacts in the tested range. Overall, among the three perturbation types, AWGN produces the most consistent severity-related decline, whereas EMG and EOG artifacts mainly induce small fluctuations around the baseline. The maintenance of accuracy near 0.95 across all noisy conditions supports the robustness of the proposed classification pipeline against both synthetic acquisition noise and representative physiological artifacts. The details of mean accuracy values and standard deviation of Figure~\ref{fig:noise}a--c could be found in Table~S18 of the Supporting Information.

\section{Methods}\label{sec:methods}

\subsection{Dataset and sample construction}\label{subsec:dataset}

In this study, we used the CHB-MIT scalp EEG dataset~\cite{PhysioNet-chbmit-1.0.0}, compiled by Boston Children's Hospital and MIT and hosted on the PhysioNet platform. This dataset is widely recognized as a standard public benchmark for epileptic seizure prediction. It comprises 23 pediatric patients, including 5 male patients aged 3 to 22 years and 18 female patients aged 1.5 to 19 years, labeled \texttt{chb01} to \texttt{chb24}. Here, \texttt{chb21} corresponds to a retest recording of \texttt{chb01} acquired 1.5 years later, yielding 23 unique subjects in total. All EEG signals were recorded in European Data Format with a uniform sampling rate of 256~Hz and 16-bit resolution. The recording channels follow the international 10--20 EEG electrode placement system. Most patient records include 23 EEG channels, and some also contain additional leads such as electrocardiography. The dataset spans approximately 983 hours and includes 198 clinically annotated seizures with expert-labeled onset and termination times.

Preictal state identification in seizure prediction was formulated as a binary classification task that distinguished between preictal and interictal states, rather than event level seizure prediction. For each patient, we randomly selected 13 nonoverlapping 10-s EEG windows and each window contained the same 18 EEG channels, yielding 234 channel-level nodes per class. Each channel-specific time series was treated as one classification unit and inherited the label of its parent window. Hence the dataset per patient has 468 channel-level nodes with 234 preictal states and 234 interictal states after signal preprocessing, whose full information could be found in Section~S1 of the Supporting Information. The preictal state was defined as EEG segments within the 30-minute interval prior to seizure onset. The interictal definition was more rigorous. Interictal segments were required to be at least three hours from any seizure onset and at least one hour after seizure termination to avoid contamination by postictal activity. Patient \texttt{chb24} was excluded because of inconsistencies between the EDF+ onset timestamps and the channel configuration. To maintain inter-patient dimensional consistency, the 18 standard channels shared by all 23 patients were retained in all experiments.

\subsection{Transductive evaluation protocol and nested cross-validation}
The proposed framework treats all EEG channels from an individual patient as nodes in a fixed patient-specific correlation network. The network was therefore constructed from all available channels before downstream CV. Only the EEG signals and their pairwise Pearson correlations were used during network construction and feature generation. The preictal and interictal labels were not used in network filtration, persistent-Laplacian analysis, chaotic dynamical embedding, or node-wise differentiation. CV was subsequently performed at the node-label classification stage. In each fold, the labels of the held-out nodes were masked, and only the representations and labels of the training nodes were used to optimize and fit the classifier. The resulting evaluation therefore represents a patient-specific transductive node classification setting in which the complete unlabeled graph structure is observable, whereas the labels of the held-out nodes remain unknown.

For each patient, model evaluation was performed using stratified nested CV. The outer loop comprised five folds and was used exclusively for performance evaluation. Within each outer-training partition, stratified three-fold CV jointly selected among seven fingerprint configurations (T, G, D, T+G, T+D, G+D, and ALL), five classifiers (LR, SVM, RF, KNN, and GBDT), and their hyperparameters. Feature normalization was fitted within each inner-training fold. The selected configuration was subsequently refitted using the complete outer-training partition and evaluated once on the corresponding outer-test partition. Outer-test labels and performance were unavailable throughout model selection. Patient-level metrics were averaged across the five outer folds, and cohort-level results were summarized across the 23 independently evaluated patients using the mean, standard deviation, and 95\% bootstrap confidence interval.

CV partitions were defined at the channel-node level and were not constructed by seizure event or chronological order. Consequently, the evaluation measures transductive classification of held-out node labels within the fixed network, rather than generalization to unseen windows or seizures. Additionally, all graph construction, feature extraction, normalization, hyperparameter selection, and classifier evaluation procedures were conducted separately for each patient. Nodes from different patients were not pooled to train a common classifier, and no patient was treated as an external test domain. Hence, the cohort level metrics are arithmetic summaries of 23 independent within-patient evaluations and do not represent leave-one-patient-out or cross-patient generalization performance.

To assess sensitivity to window selection, we repeated the complete evaluation for ten independently sampled window sets using ten different selection seeds. In each replicate, 13 preictal and 13 interictal windows were selected per patient under the original eligibility criteria. We reconstructed the correlation network and ten functional subgraphs, recomputed all three fingerprint families and node-removal differentiation, and reran the complete five-outer-fold, three-inner-fold nested-CV procedure. The fold-assignment scheme, hyperparameter search spaces, and all other settings were held fixed, so that only window selection varied. For each patient and replicate, accuracy was averaged across the five outer test folds. We then summarized accuracy across replicates for each patient and across patients for each replicate.

\subsection{Correlation-based network construction and interaction mapping}\label{subsec:feature_pipeline}

After preprocessing, each sample is defined as an EEG channel $x_i \in \mathbb{R}^{T}$. For a given patient, a network of $N$ nodes can be constructed. The associated node set is denoted by $\mathcal{V}=\{1,\dots,N\}$ and each $x_i$ was assigned to one node. The edge weight between nodes $i$ and $j$ was defined by the Pearson correlation coefficient:
\begin{equation}
P_{ij} =
\frac{\sum_{t=1}^{T}\left(x_i(t)-\bar{x}_i\right)\left(x_j(t)-\bar{x}_j\right)}
{\sqrt{\sum_{t=1}^{T}\left(x_i(t)-\bar{x}_i\right)^2}\sqrt{\sum_{t=1}^{T}\left(x_j(t)-\bar{x}_j\right)^2}}.
\end{equation}
To focus on positively correlated similarity structures, negative correlation edges were thresholded to zero.
Topological relations or connectivities among network nodes are basic ingredients in network nonlinear dynamical models. The connectivity matrix should agree with the driven and response relation between two dynamics systems, so we assume that all nodes in the network are mutually linked and their interactions increase as a function of their correlation $\textbf{A}_{ij}(P_{ij})$. The simplest form of the connectivity matrix is the Kirchoff (or connectivity) matrix generated by cut-off correlation coefficient $\sigma_{ij}$,
\begin{equation}
    A_{ij} = 
\begin{cases} 
1, & \forall P_{ij} \leq \sigma_{ij}, \, i \neq j \\
0, & \forall P_{ij} > \sigma_{ij}, \, i \neq j. \\
-\sum_{j \neq i} A_{ij}, & \forall i = j
\end{cases}\label{aij1}
\end{equation}
In order to take the correlation effect into a more practical way in the consideration, the correlation was subsequently rescaled through a generalized exponential function:
\begin{equation}
    A_{ij} = 
\begin{cases} 
e^{(P_{ij}-1)^k / k\sigma_{ij}^k}, & \forall i \neq j, k = 1, 2, \cdots \\
-\sum_{j \neq i} A_{ij}, & \forall i = j,
\end{cases}\label{aij2}
\end{equation}
where $\sigma_{ij}$ is a tuneable parameter and the characteristic correlation between nodes in the present work. Equations \ref{aij1}–\ref{aij2} can be utilized to convert the geometric properties of a network into topological connections or connectivity patterns. The connectivity matrix $\textbf{A}$ is an $N\times N$ symmetric, diagonally dominant matrix, and its entries do not represent interaction potentials between nodes. For simplicity, we assume that the characteristic correlations for all nodes are identical, so $\sigma_{ij} = \sigma$.

\subsection{Coupled nonlinear-dynamics fingerprints (\DynFPs{})}\label{subsec:DyFps} 
As a brain neural network consists of $N$ nodes and exhibits spatiotemporal complexity in $\mathbb{R}^{3N} \times \mathbb{R}^{+}$, its dynamical behavior can be modeled through network nonlinear dynamics composed of $N$ interacting nonlinear oscillators in $\mathbb{R}^{nN} \times \mathbb{R}^{+}$, where $n$ denotes the dimensionality of each individual oscillator. Accordingly, we consider an $n \times N$-dimensional nonlinear system governed by $N$ coupled nonlinear oscillators:
\begin{equation}
    \frac{d\mathbf{u}}{dt} = \mathbf{F}(\mathbf{u}) + \mathbf{E}\mathbf{u},
\label{nonfun}
\end{equation}
where $\mathbf{u} = (\mathbf{u}_1, \mathbf{u}_2, \dots, \mathbf{u}_N)^T$ denotes the array of state functions for the $N$ nonlinear oscillators, and $\mathbf{u}_i = (u_{i1}, u_{i2}, \dots, u_{in})^T$ is the $n$-dimensional nonlinear function corresponding to the $i$-th oscillator. $\mathbf{F}(\mathbf{u}) = (f(\mathbf{u}_1), f(\mathbf{u}_2), \dots, f(\mathbf{u}_N))^T$ represents the array of nonlinear functions for the $N$ oscillators. The coupling matrix is defined as $\mathbf{E} = \varphi \mathbf{A} \otimes \mathbf{\Gamma}$, where $\varphi$ denotes the overall coupling strength, $\mathbf{A}$ is the $N \times N$ connectivity matrix introduced previously, and $\mathbf{\Gamma}$ is an $n \times n$ linking matrix. Although this formulation allows the incorporation of all physical interactions among neurons, the present work places greater emphasis on the geometric structure induced by inter-neuronal correlation. Consequently, the brain neural network is represented in terms of topological relationships or connectivity patterns.

In Eq.\ref{nonfun} of the proposed brain neural network model, the explicit form of the nonlinear function $\mathbf{F}(\mathbf{u})$ is not specified. From a mathematical perspective, it is well known that double-well or triple-well functions may lead to multiple local minima, giving rise to bistability or multistability, and rendering the system sensitive to small perturbations that can induce instability or chaotic behavior. For simplicity, the present study adopts a set of Lorenz attractors as a representative example \cite{lorenz1963deterministic}. Each Lorenz oscillator, defined in three dimensions with $u_i = (x_i, y_i, z_i)^T$, is described as follows:
\begin{equation}
\begin{aligned}
\dot{x_i} &= \delta(y_i - x_i),\\
\dot{y_i} &= x_i(\gamma - z_i) - y_i,\\
\dot{z_i} &= x_iy_i - \beta z_i,
\end{aligned}
\end{equation}
where $i=1,2,...,N$ and with parameters $\delta = 10$, $\gamma = 60$, and $\beta = 8/3$, the Lorenz attractor with almost all points in the phase space is created. The Lorenz equations are solved through the fourth-order Runge-Kutta scheme in the present work. After the transient stage, the final 50 steps are retained as the steady-state window. 

The coupled network dynamics is described by:
\begin{equation}
\frac{du_i}{dt} = f(u_i) + \varphi \sum_{j=1}^{N} A_{ij}\bigl(u_j - u_i\bigr), \quad i = 1, \dots, N,  
\end{equation}
where $A_{ij}$ is the adjacency connectivity matrix element and we set $\varphi=0.42$. Numerical integration is performed with a fourth-order Runge-Kutta scheme using time step $\Delta t = 0.01$ for 100 steps. The detail of parameters used in numerical simulation can be found in Table~S19 of the Supporting Information. To summarize the collective steady-state response, all node-wise $x$ trajectories in this window are pooled and flattened into a single empirical response vector. Six statistics extracted from this vector are used for dynamical fingerprints (\DynFPs{}), including the mean, median, variance, maximum, minimum, and standard deviation. 

\subsection{Persistent Laplacian-based topological and geometric fingerprints (\TopFPs{} and \GeoFPs{} )}\label{subsec:PL} 
Persistent homology (PH), a recently developed branch of algebraic topology, provides a multiscale framework for representing topological structures and has been successfully applied in diverse domains such as mathematics and chemistry \cite{su2025topological,wee2025review}. Through a filtration process, PH generates a sequence of topological spaces, enabling the analysis of data across multiple scales. However, PH is limited in its ability to characterize the homotopic evolution of shapes during the filtration process. To overcome this limitation, the persistent Laplacian (PL), also referred to as persistent spectral graph (PSG) \cite{wang2020persistent}, has been introduced. In contrast to PH, PL preserves topological invariants through its harmonic spectra while additionally encoding the homotopic geometric evolution of the data via its non-harmonic spectra. In the present work, the topological invariant and homotopic geometric evolution correspond to topological fingerprints (\TopFPs{}) and geometric fingerprints (\GeoFPs{}), respectively.

PL belongs to the broader family of persistent topological Laplacians, including the persistent Hodge Laplacian defined on Riemannian manifolds, as well as extensions on cellular sheaves, hyperdigraphs, and path complexes \cite{chen2021evolutionary}. Similar to PH, PL employs a filtration process to construct a sequence of geometric objects and their associated topological spaces. On these spaces, persistent spectral graphs are defined. The evolution of the null space dimension of the PL operator throughout the filtration reflects the persistence of topological invariants, whereas its non-zero eigenvalues and corresponding eigenfunctions describe the geometric deformation of the data along the filtration.

Let $K$ be an oriented simplicial complex with a filtration given by a nested sequence of subcomplexes:
\begin{equation}
 \emptyset=K_{0}\subseteq K_{1}\subseteq K_{2}\subseteq\cdot\cdot\cdot\subseteq K_{m}=K. \end{equation}
For each subcomplex $K_t$, denote its chain group by $C_q(K_t)$, and define the boundary operator $\partial_{q}^{t}:C_{q}(K_{t})\rightarrow C_{q-1}(K_{t})$ for $0 < q \leq \operatorname{dim} K_t$. For any $q$-simplex $\sigma_{q}=[v_{0},\cdot\cdot\cdot,v_{q}] \in K_t$ the boundary operator is expressed as
\begin{equation}
 \partial_{q}^{t}(\sigma_{q})=\sum_{i=0}^{q}(-1)^{i}\sigma_{q-1}^{i} ,
\end{equation}
where $\sigma_{q-1}^{i}=[v_{0},\cdot\cdot\cdot,\hat{v_{i}},\cdot\cdot\cdot,v_{q}]$ represents the oriented $(q-1)$-simplex formed by the exclusion of vertex $v_i$. The adjoint of the boundary operator, denoted by $(\partial_q^t)^*$, defines the corresponding coboundary operator, mapping $C_{q-1}(K_{t})$ to $C_{q}(K_{t})$ via the natural isomorphism between chain and cochain groups.

For notational simplicity, let $C_q^t = C_{q}(K_{t})$. Using the natural inclusion from $C_{q-1}^t$ to $C_{q-1}^{t+p}$, one defines the subset
\begin{equation}
    C_{q}^{t,p}=\{\beta\in C_{q}^{t+p}|\partial_{q}^{t+p}(\beta)\in C_{q-1}^{t}\}, 
\end{equation}
which consists of chains in $C_q^{t+p}$ whose boundaries lie in $C_{q-1}^{t}$. On this subset, the $p$-persistent boundary operator is defined as $\mathfrak{\partial}_{q}^{t,p}:C_q^{t,p}\rightarrow C_{q-1}^t$, together with its adjoint operator $(\partial_q^{t,p})^*$. Based on these operators, the $q$-th order $p$-persistent Laplacian is defined as
\begin{equation}
    \Delta_{q}^{t,p}=\partial_{q+1}^{t,p}(\partial_{q+1}^{t,p})^{*}+(\partial_{q}^{t})^{*}{\partial}_{q}^{t}.
\end{equation}
If $C_{q}^{t,p}$ has a canonical orthonormal basis, its matrix representation is given by
\begin{equation}
    \mathcal{L}_{q}^{t,p}=\mathcal{B}_{q+1}^{t,p}(\mathcal{B}_{q+1}^{t,p})^{T}+(\mathcal{B}_{q}^{t})^{T}\mathcal{B}_{q}^{t}.
\end{equation}
The topological invariants associated with PH can be recovered from the kernel of the PL operator, namely,
\begin{equation}
    \beta_{q}^{t,p} = \text{dim ker } \partial_{q}^{t}- \text{dim im} \partial_{q+1}^{t,p} = \text{dim ker } \mathcal{L}_{q}^{t,p} ,
\end{equation}
which corresponds to the multiplicity of zero eigenvalues of $\mathcal{L}_{q}^{t,p}$ and are used for topological fingerprints (\TopFPs{}). 

In this work, the zero- and first-order PLs are considered for brain network analysis. The Betti numbers derived from the null space track the number of $q$-dimensional topological features that persist across filtration levels, providing the same invariant information as PH. In addition, the non-zero eigenvalues of PL encode the homotopic geometric evolution of the data during filtration. The statistical features of PL non-zero eigenvalues are used for geometric fingerprints (\GeoFPs{}).

It is worth emphasizing that, instead of applying conventional filtration directly on simplicial complexes as in standard PH or PL approaches, a Laplacian-based filtration is adopted in this study. Specifically, a weighted Laplacian matrix is used to construct a sequence of subgraphs through an accumulation threshold. The Laplacian matrix is defined as 
\begin{equation}
    L = (l_{ij}), \quad l_{ii} = -\sum_{j=1}^n l_{ij},
\end{equation}
and for $i \neq j$, let $l_{\text{max}} = \text{max}(l_{ij})$, $l_{\text{min}}=\text{min}(l_{ij})$, and $d=l_{\text{max}}-l_{\text{min}}$. The $k$-th filtered Laplacian is then constructed as 
\begin{equation}
    l_{ij}^k =
\begin{cases}
0, & l_{ij} \leq \dfrac{k}{p} d + l_{\min}, \\
-1, & \text{otherwise},
\end{cases}
\quad
l_{ii}^k = -\sum_{j=1}^n l_{ij}^k.
\end{equation}
This procedure generates a sequence of subgraphs encoding interaction strengths at different levels, enabling multiscale topological and geometric analysis.

\section{Conclusion}
The proposed chaotic dynamics–regulated topological learning (CDRTL) framework integrates nonlinear dynamics, topology, and geometry for offline, patient-specific discrimination of preictal and interictal EEG channel-level nodes within a fixed transductive network. Evaluations on the CHB-MIT dataset demonstrate high patient-specific accuracy, robust performance under consensus aggregation, and strong resilience to noise perturbations. Notably, dynamical features play a central role. The sensitivity of chaotic systems to initial conditions may amplify subtle preictal variations into highly discriminative representations, while topological and geometric features provide complementary structural information. But their incremental predictive value relative to their computational cost remains to be established. The framework further shows consistent performance across different dynamical systems, indicating that its effectiveness is not tied to a specific model. Together, these components establish a unified and expressive characterization of complex neural states.

Several limitations remain. The current evaluation is limited to balanced, patient-specific preictal–interictal EEG channel classification and does not include seizure-wise continuous testing. Consequently, event level sensitivity and false alarms per hour were not evaluated, leaving cross-subject generalizability and broader clinical applicability to be further explored, which are important for evaluating a clinically deployable seizure-warning system. Additionally, the framework is designed for offline analysis, and real-time deployment requires further development. Future work will develop an inductive representation that enables rigorous leave-one-patient-out evaluation and prospective testing on continuous EEG recordings. Further extensions may also incorporate more advanced topological Laplacian formulations to enhance representation capacity \cite{wei2025persistent}. Moreover, computational runtime and memory requirements were not benchmarked. Future evaluations should quantify the additional cost and predictive benefit of the spectral fingerprints relative to dynamical feature-only implementation, as well as assess online scalability after an inductive extension becomes available.

\section*{Data and code availability}
The code and associated data used in this study are available at https://github.com/Wzhgeek/CDRTL.git.

\section*{Acknowledgments}
This work was supported in part by Wuhan Textile University. 
The work of M.Z., X.W. and XXW was supported in part by MSU Research Foundation. 

\section*{Competing interests}
The authors declare no competing interests.

\bibliographystyle{abbrv}
\bibliography{ref}

\end{document}